\documentclass[12pt]{article}
\usepackage{graphicx}
\usepackage{subfig}
\usepackage{epsfig}
\usepackage{rotating}
\usepackage{amsmath}
\usepackage{multirow}
\def\br(#1,#2){\left\langle#1#2\right\rangle}
\def\sq(#1,#2){\left[#1#2\right]}
\def\s(#1,#2){s_{#1 #2}}
\def\t(#1,#2,#3){s_{#1 #2 #3}}

\begin{document}

\begin{titlepage}
%\hspace*{\fill}\parbox[t]{5cm}
%{\today} \vskip2cm
\begin{flushright}
	{\today} \\
	MPP-2013-153 \\
	CP3-13-29
\end{flushright}
\vskip2cm
\begin{center}
{\Large Constraints on electroweak effective operators at one loop} \\
\medskip
\bigskip\bigskip\bigskip\bigskip
{\bf Harrison Mebane,$^1$ Nicolas Greiner,$^{1,2}$ Cen Zhang,$^{1,3}$ and Scott Willenbrock$^1$}\\
\bigskip\bigskip\bigskip
$^1$Department of Physics, University of Illinois at Urbana-Champaign \\ 1110 West Green Street, Urbana, Illinois 61801, USA \\
\bigskip
$^2$Max-Planck-Institut f\"ur Physik, F\"ohringer Ring 6, 8085 M\"unchen, Germany \\
\bigskip
$^3$Centre for Cosmology, Particle Physics and Phenomenology (CP3) \\
Universit\'e Catholique de Louvain, B-1348 Louvain-la-Neuve, Belgium
\end{center}

\bigskip\bigskip\bigskip

\begin{abstract}

We derive bounds on nine dimension-six operators involving electroweak gauge bosons and the Higgs boson from precision electroweak data.  Four of these operators contribute at tree level, and five contribute only at one loop.  Using the full power of effective field theory, we show that the bounds on the five loop-level operators are much weaker than previously claimed, and thus much weaker than bounds from tree-level processes at high-energy colliders.

\end{abstract}

\end{titlepage}

\section{Introduction}
\label{sec:intro}

The discovery of the Higgs boson at the LHC finally completes the Standard Model.
The next step is the discovery of physics beyond the Standard Model.  This can be done directly by searching for new particles, or indirectly by searching for new interactions
of the Standard Model particles.  Indirect searches for new physics can be done model-independently by means of effective field theory \cite{Weinberg:1978kz,Buchmuller:1985jz,Grinstein:1991cd}.

An effective field theory is a low-energy approximation of a higher-energy theory.  By integrating out high-energy degrees of freedom, one obtains a low-energy theory that includes additional effective interactions which involve only low-energy fields.  One obtains a perturbative expansion in which effective interactions, or operators, are suppressed by inverse powers of the mass scale of the physics which has been integrated out.  If, as in our case, one does not know the high-energy theory, a complete operator basis can be written down at each order.

The Standard Model operators have mass dimension four or less.  The only possible operator of dimension five generates Majorana neutrino masses and does not concern us here
\cite{Weinberg:1979sa}. Thus, the lowest-dimension effective operators are of dimension six.  We can write down an effective field theory which extends the Standard Model in the following form
\begin{equation}
	\mathcal{L}_{eff} = \mathcal{L}_{SM} + \sum_i \frac{c_i}{\Lambda^2}\mathcal{O}_i + \ldots
\end{equation}
where the $\mathcal{O}_i$ are dimension-six operators, $\Lambda$ is the mass scale of new physics, and the $c_i$ are dimensionless coefficients that reflect our ignorance of the high-energy theory.  This expansion reduces to the Standard Model in the limit $\Lambda \rightarrow \infty$.  A complete basis of operators $\mathcal{O}_i$ comprises operators which are independent with respect to equations of motion and which are $SU(3)\times SU(2)\times U(1)$ gauge-invariant \cite{Buchmuller:1985jz,Grzadkowski:2010es}.  Aside from reducing the number of independent operators, this latter condition guarantees a consistent framework for performing loop calculations.  That is, divergences produced by an operator at a given order in $1/\Lambda$ can always be absorbed by other operators at the same order in $1/\Lambda$.  Thus the renormalization program can be carried out, order-by-order, in any complete effective field theory.

In this paper we use the precision electroweak data in Table~\ref{tab:observables} to calculate bounds on nine dimension-six operators containing only gauge boson fields and Higgs doublets.   All contributions from the nine operators can be represented as gauge boson self-energies, also called oblique corrections \cite{Kennedy:1988sn,Peskin:1991sw}.  Five of the operators contribute only at one loop; the four remaining operators contribute at tree level and must be included in order to absorb one-loop ultraviolet divergences from the other five operators.

\begin{table}[t]\footnotesize
\begin{tabular}{|c|c|c|}
\hline
& Notation & Measurement \\
\hline
$Z$-pole & $\Gamma_Z$ & Total $Z$ width \\
& $\sigma_{\rm had}$ &	Hadronic cross section \\
& $R_f$($f=e,\mu,\tau,b,c$) & Ratios of decay rates \\
& $A_{FB}^{0,f}$($f=e,\mu,\tau,b,c,s$) & Forward-backward asymmetries \\
& $\bar{s}_l^2$ & Hadronic charge asymmetry \\
& $A_f$($f=e,\mu,\tau,b,c,s$) & Polarized asymmetries \\
\hline
Fermion pair & $\sigma_f$($f=q,e,\mu,\tau)$ & Total cross sections for $e^+e^-\rightarrow f\bar{f}$ \\
production at LEP2 & $A_{FB}^f$($f=\mu,\tau$) & Forward-backward asymmetries for $e^+e^-\rightarrow f\bar{f}$ \\
\hline
$W$ mass & $m_W$ & $W$ mass from LEP and Tevatron \\
and decay rate & $\Gamma_W$ & $W$ width from Tevatron \\
\hline
DIS & $Q_W(Cs)$	& Weak charge in Cs \\
and & $Q_W(Tl)$ & Weak charge in Tl \\
atomic parity violation	& $Q_W(e)$ & Weak charge of the electron \\
& $g_L^2,g_R^2$	& $\nu_\mu$-nucleon scattering from NuTeV \\
& $g_V^{\nu e},g_A^{\nu e}$ & $\nu$-$e$ scattering from CHARM II \\
\hline
\end{tabular}
\caption{Precision electroweak quantities.  Data taken from \cite{Beringer:1900zz,Schael:2013ita}.\label{tab:observables}}
\end{table}

Similar analyses have been done previously \cite{De Rujula:1991se,Hagiwara:1992eh,Hagiwara:1993ck,Alam:1997nk}.  These previous analyses did not appreciate that unambiguous bounds can be obtained on the five loop-level operators.\footnote{A similar calculation, with the same shortcomings, is performed for a model with no Higgs field in Ref.~\cite{Burgess:1993qk}.}   We recently showed that the bounds on two of these five operators are much weaker than had been obtained in previous analyses \cite{Mebane:2013cra}.  In this paper we extend this analysis to all five of the loop-level operators.

Because precision electroweak data are taken at ``low'' energies, around the Z boson mass or below, there will often be significant suppression of operator contributions, of the order $\hat{s}/\Lambda^2$, where $\hat{s}$ is the usual Mandelstam variable.  Furthermore, the five operators contributing only at one loop receive an additional suppression of $1/(4\pi)^2$.  It is therefore reasonable to ask what advantages precision measurements offer.  For one, electroweak data is known to far greater precision than high-energy collider data from the Tevatron and LHC.  In addition, the effective operator contribution is not always energy-dependent; it is often proportional to $v^2/\Lambda^2$.  In this case, there is no disadvantage to using low-energy data.\footnote{Here we are considering only interference terms between the effective operators and the Standard Model.  If $c_i^2/\Lambda^4$ terms are included, there will in general be energy dependence.}  We therefore perform this analysis both as an illustration of the power of effective field theory and in order to compare our loop-level results with tree-level results from high-energy colliders.

In Section~\ref{sec:operators}, we discuss the nine effective operators to be examined in this paper.  In Section~\ref{sec:corrections}, we outline the framework for computing the effect of oblique corrections on electroweak observables.  We present bounds on the effective operators in Section~\ref{sec:results}, and conclude in Section~\ref{sec:conclusions}.

\section{Electroweak Effective Operators}
\label{sec:operators}

In this paper, we are interested in the effects of new physics on precision electroweak data.  Here we examine the set of operators that involves only gauge and Higgs bosons.    Five of these contribute only at one loop \cite{Hagiwara:1993ck}:
\begin{subequations}
\begin{align}
	\mathcal{O}_{WWW} & = \text{Tr}\:\hat{W}^{\mu}_{\phantom{\mu}\nu} \hat{W}^{\nu}_{\phantom{\nu}\rho}
	\hat{W}^{\rho}_{\phantom{\rho}\mu} \label{abc} \\
	\mathcal{O}_{W} & = \left(D_\mu \phi\right)^\dagger \hat{W}^{\mu\nu} \left(D_\nu \phi\right) \\
	\mathcal{O}_{B} & = \left(D_\mu \phi\right)^\dagger \hat{B}^{\mu\nu} \left(D_\nu \phi\right) \label{def} \\
	\mathcal{O}_{WW} & = \phi^\dagger \hat{W}^{\mu\nu} \hat{W}_{\mu\nu} \phi \\
	\mathcal{O}_{BB} & = \phi^\dagger \hat{B}^{\mu\nu} \hat{B}_{\mu\nu} \phi
\end{align}
\label{loopops}
\end{subequations}
where $\hat{B}_{\mu\nu} = ig^\prime\frac{1}{2} B_{\mu\nu}$, $\hat{W}_{\mu\nu} = ig\frac{\sigma^{a}}{2} W^{a}_{\mu\nu}$ and $\sigma^a$ is the $a$th Pauli matrix.  The covariant derivative is defined as
\begin{equation}
	D_\mu \phi = \left(\partial_\mu - ig^\prime\frac{1}{2} B_\mu - ig\frac{\sigma^{a}}{2} W^{a}_\mu\right)\phi
\end{equation}
Table \ref{tab:diagrams} lists all one-loop Feynman graphs and the operators that contribute to them.  The above operators affect precision electroweak observables in two different ways.  All five operators affect gauge boson self-energies through loop corrections.  In addition, the first three operators alter the fermion-fermion-boson vertices.  It would seem as if the final two operators, $\mathcal{O}_{BB}$ and $\mathcal{O}_{WW}$, contribute to gauge boson self-energies at tree level when the Higgs doublets take their vacuum expectation values; however, these contributions can be absorbed into the Standard Model gauge kinetic terms with field and coupling redefinitions.  These operators therefore only affect diagrams involving Higgs bosons \cite{Mebane:2013cra}.

\begin{table}[ht]
\centering
\begin{tabular}{|c|c|c|}
\hline
\multirow{4}{*}{\includegraphics[width=1in]{vert}}
& $\delta\Gamma_W$ & $\mathcal{O}_{WWW}$, $\mathcal{O}_B$, $\mathcal{O}_W$ \\
& $\delta\Gamma_Z$ & $\mathcal{O}_{WWW}$, $\mathcal{O}_B$, $\mathcal{O}_W$ \\
& $\delta\Gamma_\gamma$ & $\mathcal{O}_{WWW}$, $\mathcal{O}_B$, $\mathcal{O}_W$ \\
& & \\ \hline
\multirow{2}{*}{\includegraphics[width=1in]{loop1}} % HISZ diagram a
& $\Pi_{WW}$ & $\mathcal{O}_{WWW}$, $\mathcal{O}_B$, $\mathcal{O}_W$ \\
& $\Pi_{ZZ}$ & $\mathcal{O}_{WWW}$, $\mathcal{O}_B$, $\mathcal{O}_W$ \\
\multirow{3}{*}{\includegraphics[width=1in]{loop2}} % HISZ diagram a
& $\Pi_{\gamma\gamma}$ & $\mathcal{O}_{WWW}$, $\mathcal{O}_B$, $\mathcal{O}_W$ \\
& $\Pi_{\gamma Z}$ & $\mathcal{O}_{WWW}$, $\mathcal{O}_B$, $\mathcal{O}_W$ \\
& & \\ \hline
\multirow{2}{*}{\includegraphics[width=1in]{loop3}} % HISZ diagram b,d
& $\Pi_{WW}$ & $\mathcal{O}_B$, $\mathcal{O}_W$, $\mathcal{O}_{WW}$ \\
& $\Pi_{ZZ}$ & $\mathcal{O}_B$, $\mathcal{O}_W$, $\mathcal{O}_{BB}$, $\mathcal{O}_{WW}$ \\
\multirow{3}{*}{\includegraphics[width=1in]{loop4}} % HISZ diagram b,d
& $\Pi_{\gamma\gamma}$ & $\mathcal{O}_B$, $\mathcal{O}_W$ \\
& $\Pi_{\gamma Z}$ & $\mathcal{O}_B$, $\mathcal{O}_W$, $\mathcal{O}_{BB}^*$, $\mathcal{O}_{WW}^*$
\\ & & $^*$ top diagram only \\ \hline
\multirow{2}{*}{\includegraphics[width=1in]{loop5}} % HISZ diagram c
& $\Pi_{WW}$ & $\mathcal{O}_W$\\
& $\Pi_{ZZ}$ & $\mathcal{O}_B$, $\mathcal{O}_W$ \\
\multirow{3}{*}{\includegraphics[width=1in]{loop6}} % HISZ diagram c
& $\Pi_{\gamma\gamma}$ & $\mathcal{O}_B$, $\mathcal{O}_W$ \\
& $\Pi_{\gamma Z}$ & $\mathcal{O}_B$, $\mathcal{O}_W$ \\ & & \\ \hline
\multirow{4}{*}{\includegraphics[width=1in]{loop11}} % HISZ first diagram g
& $\Pi_{WW}$ & $\mathcal{O}_{BB}$, $\mathcal{O}_{WW}$ \\
& $\Pi_{ZZ}$ & $\mathcal{O}_{BB}$, $\mathcal{O}_{WW}$ \\
& $\Pi_{\gamma\gamma}$ & \\
& $\Pi_{\gamma Z}$ & \\ \hline
\multirow{4}{*}{\includegraphics[width=1in]{loop7}} % HISZ second diagram f
& $\Pi_{WW}$ & $\mathcal{O}_{WW}$ \\
& $\Pi_{ZZ}$ & $\mathcal{O}_{BB}$, $\mathcal{O}_{WW}$ \\
& $\Pi_{\gamma\gamma}$ & $\mathcal{O}_{BB}$, $\mathcal{O}_{WW}$ \\
& $\Pi_{\gamma Z}$ & $\mathcal{O}_{BB}$, $\mathcal{O}_{WW}$ \\ \hline
\multirow{4}{*}{\includegraphics[width=1in]{loop9}} % HISZ first diagram f
& $\Pi_{WW}$ & $\mathcal{O}_{WW}$ \\
& $\Pi_{ZZ}$ & $\mathcal{O}_{BB}$, $\mathcal{O}_{WW}$ \\
& $\Pi_{\gamma\gamma}$ & $\mathcal{O}_{BB}$, $\mathcal{O}_{WW}$ \\
& $\Pi_{\gamma Z}$ & $\mathcal{O}_{BB}$, $\mathcal{O}_{WW}$ \\ \hline \multirow{4}{*}{\includegraphics[width=1in]{loop10}} % HISZ third diagram f
& $\Pi_{WW}$ & $\mathcal{O}_{WW}$ \\
& $\Pi_{ZZ}$ & $\mathcal{O}_{BB}$, $\mathcal{O}_{WW}$ \\
& $\Pi_{\gamma\gamma}$ & $\mathcal{O}_{BB}$, $\mathcal{O}_{WW}$ \\
& $\Pi_{\gamma Z}$ & $\mathcal{O}_{BB}$, $\mathcal{O}_{WW}$ \\ \hline \multirow{4}{*}{\includegraphics[width=1in]{loop12}} % HISZ second diagram e
& $\Pi_{WW}$ & $\mathcal{O}_{WW}$ \\
& $\Pi_{ZZ}$ & $\mathcal{O}_{BB}$, $\mathcal{O}_{WW}$ \\
& $\Pi_{\gamma\gamma}$ & $\mathcal{O}_{BB}$, $\mathcal{O}_{WW}$ \\
& $\Pi_{\gamma Z}$ & $\mathcal{O}_{BB}$, $\mathcal{O}_{WW}$ \\ \hline
\multirow{4}{*}{\includegraphics[width=1in]{loop13}} % HISZ first diagram e
& $\Pi_{WW}$ & $\mathcal{O}_W$ \\
& $\Pi_{ZZ}$ & $\mathcal{O}_W$ \\
& $\Pi_{\gamma\gamma}$ & \\
& $\Pi_{\gamma Z}$ & $\mathcal{O}_W$ \\ \hline
\end{tabular}
\caption{Feynman Diagrams}
\label{tab:diagrams}
\end{table}

The one-loop self-energies above contain ultraviolet divergences.  The following set of four operators, all of which affect self-energies at tree level, is sufficient to absorb all divergences from the operators of Eq.~(\ref{loopops}) \cite{Hagiwara:1993ck}
\begin{subequations}
\begin{align}
	\mathcal{O}_{BW} & = \phi^\dagger \hat{B}^{\mu\nu} \hat{W}_{\mu\nu} \phi \\
	\mathcal{O}_{\phi,1} & = \left(D_\mu \phi\right)^\dagger\! \phi\: \phi^\dagger\! \left(D^\mu \phi\right) \\
	\mathcal{O}_{DW} & = \textrm{Tr}\:[D^\mu, \hat{W}^{\nu\rho}] [D_\mu, \hat{W}_{\nu\rho}] \\
	\mathcal{O}_{DB} & = 2\: \partial^\mu \hat{B}^{\nu\rho} \partial_\mu \hat{B}_{\nu\rho}
\end{align}
\label{treeops}
\end{subequations}

\section{One-Loop Bounds from Precision Electroweak Data}\label{sec:corrections}

The operators of Eq.~\eqref{loopops} affect the precision data only through gauge boson self-energies and fermion-fermion-boson vertices.  Table \ref{tab:diagrams} shows the diagrams which contribute.  The vertex corrections and self-energies always contribute to observables in the same gauge-invariant combinations \cite{Hagiwara:1993ck}
\begin{align}
	\label{piww}
	\overline{\Pi}_{WW} & = \Pi_{WW} + 2(q^2-m_W^2)\delta\Gamma^W \\
	\overline{\Pi}_{ZZ} & = \Pi_{ZZ} + 2c\:(q^2-m_Z^2)\delta\Gamma^Z \\
	\overline{\Pi}_{\gamma\gamma} & = \Pi_{\gamma\gamma} + 2s\:q^2\delta\Gamma^\gamma \\
	\label{piaz}
	\overline{\Pi}_{\gamma Z} & = \Pi_{\gamma Z} + s\: q^2\delta\Gamma^Z + c\:(q^2-m_Z^2) \delta\Gamma^\gamma
\end{align}
where the $\Pi_{XY}$ are the transverse parts of the gauge boson self-energies, $s$ and $c$ are the sine and cosine of the weak mixing angle, and the $\delta\Gamma^i$ are the fermion-fermion-boson vertex corrections, defined as
\begin{align}
	\delta\Gamma_\mu^{Vf\bar{f}} & = g I_3^f \gamma_\mu \frac{1}{2}(1-\gamma_5) \delta\Gamma^V \\
	\delta\Gamma_\mu^{Wf_1 f_2} & = \frac{g}{\sqrt{2}}\gamma_\mu \frac{1}{2}(1-\gamma_5) \delta\Gamma^W
\end{align}
where $V$ denotes a neutral vector boson, and $I_3^f$ denotes the third component of the fermion's isospin.

Modified self-energies contribute to precision electroweak data through corrections to the input variables $\alpha$, $m_Z$, and $s_W^2$.  The correction to $\alpha$ depends upon the type of vertex; these corrections will be labeled $\delta\alpha_\gamma$, $\delta\alpha_Z$, or $\delta\alpha_W$, depending on the mediating boson.  The modified self-energy between bosons $X$ and $Y$ is denoted $\Pi_{XY}$ in the expressions below:
\begin{align}
	\alpha + \delta\alpha_\gamma & = \alpha\left(1+\overline{\Pi}_{\gamma\gamma}^\prime(q^2)
		-\overline{\Pi}_{\gamma\gamma}^\prime(0)\right) \\
	\alpha + \delta\alpha_Z & = \alpha\left(1+\overline{\Pi}_{\gamma\gamma}^\prime(q^2)
		-\overline{\Pi}_{\gamma\gamma}^\prime(0)\right) \\
	& \qquad \qquad \times \left(1+\frac{d}{dq^2}\overline{\Pi}_{ZZ}(m_Z^2)-\overline{\Pi}_{\gamma\gamma}^\prime(q^2)
		-\frac{c^2-s^2}{c s}\overline{\Pi}_{\gamma Z}^\prime(q^2)\right) \nonumber \\
	\alpha + \delta\alpha_W & = \alpha\left(1+\overline{\Pi}_{\gamma\gamma}^\prime(q^2)
		-\overline{\Pi}_{\gamma\gamma}^\prime(0)\right) \\
	& \qquad \qquad \times \left(1+\frac{d}{dq^2}\overline{\Pi}_{WW}(m_W^2)-\overline{\Pi}_{\gamma\gamma}^\prime(q^2)
		-\frac{c}{s}\overline{\Pi}_{\gamma Z}^\prime(q^2)\right) \nonumber \\
	m_Z^2 + \delta m_Z^2 & = m_Z^2 - \overline{\Pi}_{ZZ}(m_Z^2) + \overline{\Pi}_{ZZ}(q^2)
		- (q^2-m_Z^2)\frac{d}{dq^2}\overline{\Pi}_{ZZ}(m_Z^2) \\
	s_W^2 + \delta s_W^2 & = s^2\left[1-\frac{c}{s}\overline{\Pi}_{\gamma Z}^\prime(q^2) - \frac{c^2}{c^2-s^2}
		\left(\overline{\Pi}_{\gamma\gamma}^\prime(0) + \frac{1}{m_W^2}\overline{\Pi}_{WW}(0)-\frac{1}{m_Z^2}\overline{\Pi}_{ZZ}(m_Z^2)\right)\right]
\end{align}
where $\overline{\Pi}_{XY}^\prime(q^2) = (\overline{\Pi}_{XY}(q^2)-\overline{\Pi}_{XY}(0))/q^2$ (with $\overline{\Pi}_{XY}^\prime(0) = \frac{d}{dq^2}\overline{\Pi}_{XY}(0)$).

The correction to any electroweak observable $X$ measured at an energy at or above the $Z$-pole is given by
\begin{equation}
	\delta X = \frac{\delta X}{\delta \alpha} \delta\alpha + \frac{\delta X}{\delta m_Z^2} \delta m_Z^2 +
		\frac{\delta X}{\delta s_W^2} \delta s_W^2
	\label{dx1}
\end{equation}
Low-energy observables are affected by corrections to $s_W^2$ and by changes to the $\rho$ parameter
\begin{equation}
	\delta X = \frac{\delta X}{\delta s_W^2} \delta s_W^2 + \frac{\delta X}{\delta \rho} \delta \rho
	\label{dx2}
\end{equation}
where $\delta\rho = \frac{1}{m_W^2}\overline{\Pi}_{WW}(0) - \frac{1}{m_Z^2}\overline{\Pi}_{ZZ}(0)$.

\section{Renormalization of Tree-Level Operators}
\label{sec:tree_renorm}

All divergences generated by the one-loop contributions of the operators in Eq.~\eqref{loopops} can be removed by a suitable renormalization of the coefficients $c_{BW}$, $c_{\phi}^{(3)}$, $c_{DW}$, and $c_{DB}$.
The renormalized tree-level coefficients, in the $\overline{\rm MS}$ scheme, are
\begin{align}
	c_{\phi,1}(\mu) & =
	c_{\phi,1}^0\mu^{-2\epsilon} + \frac{3g^4s^2}{128\pi^2m_W^2 c^2}\left((m_h^2 + 3m_W^2)c_B + 3m_W^2 c_W\right)\left(\frac{1}{\epsilon}-\gamma+\ln 4\pi\right) \\
	c_{DB}(\mu) & = c_{DB}^0\mu^{2\epsilon} - \frac{c_B}{192\pi^2}\left(\frac{1}{\epsilon}-\gamma+\ln 4\pi\right) \\
	c_{DW}(\mu) & = c_{DW}^0\mu^{2\epsilon} - \frac{c_W}{192\pi^2}\left(\frac{1}{\epsilon}-\gamma+\ln 4\pi\right) \\
	c_{BW}(\mu) & = c_{BW}^0 - \frac{g^2}{16\pi^2}\left(c_{WW} + \frac{s^2}{c^2} c_{BB} - \frac{3}{2} c_{WWW}g^2
	- \frac{1}{24c^2m_W^2}c_B(3c^2m_h^2 + (7+20c^2)m_W^2) \right. \\
	& \hspace{2.3in} \left. -	\frac{1}{24c^2m_W^2}c_W(3c^2m_h^2 - (3+12c^2)m_W^2)\right)\left(\frac{1}{\epsilon}-\gamma+\ln 4\pi\right) \nonumber
\end{align}
where the superscript ``0'' indicates the bare coefficient.

\section{Results}
\label{sec:results}

We now take all of the self-energy corrections from Appendix~\ref{apdx:se} and compute oblique corrections to the precision electroweak observables listed in Table~\ref{tab:observables}.  We use the following values for input parameters
\begin{align}
	\alpha(m_Z) = 1/128.91, \qquad v & = 246.2\textrm{ GeV}, \qquad m_Z = 91.1876\textrm{ GeV}, \qquad
	m_h = 125\textrm{ GeV} \\
	m_t = 172.9\textrm{ GeV},& \qquad m_b = 4.79\textrm{ GeV}, \qquad m_\tau = 1.777\textrm{ GeV} \nonumber
\end{align}
The masses of all other fermions are neglected.  We set the renormalization scale to $\mu = M_Z$ in the tree-level coefficients.

We use the $\chi^2$ statistic to compute bounds on the operators.
\begin{equation}
	\chi^2 = \sum_{i,j} \chi^i \left(\sigma^{-1}\right)_{ij} \chi^j
\end{equation}
where $\sigma_{ij}$ is the error matrix, and
\begin{equation}
	\chi^i = \left(X^i_{SM}-X^i_{exp}+ \sum_k\frac{c_k}{\Lambda^2}X^i_k\right)
\end{equation}
where the sum on $k$ runs over all loop- and tree-level operators.

We begin by writing $\chi^2$ in the following way
\begin{equation}
	\chi^2 = \chi^2_{min} + \frac{\sum_{ij}(c_i-\hat{c}_i)M_{ij}(c_j-\hat{c}_j)}{\Lambda^4}
\end{equation}
where the $i,j$ sum is over all nine operators.  The $\hat{c}_i$ are best-fit values.  We then arrive at $1\sigma$ bounds by solving the equation
\begin{equation}
	\frac{\sum_{ij}(c_i-\hat{c}_i)M_{ij}(c_j-\hat{c}_j)}{\Lambda^4} = 1
\end{equation}
It is cleanest to diagonalize the matrix $M$ and present bounds on the nine linearly independent combinations of operators.  Those bounds appear below
\begin{align}
	& \begin{pmatrix}
		-0.164 & 0.986 & -0.018 & 0.025 & -0.000 & -0.001 & 0.000 & -0.000 & -0.000 \\
		-0.494 & -0.103 & -0.832 & -0.230 & -0.000 & -0.002 & 0.001 & 0.000 & -0.000 \\
		-0.838 & -0.131 & 0.527 & -0.051 & 0.000 & -0.001 & 0.001 & -0.001 & -0.000 \\
		-0.165 & -0.006 & -0.170 & 0.972 & -0.001 & -0.000 & 0.002 & 0.000 & 0.000 \\
		0.001 & -0.000 & 0.001 & -0.001 & -0.913 & -0.218 & 0.145 & -0.312 & 0.011 \\
		-0.002 & 0.000 & -0.001 & -0.001 & -0.156 & 0.961 & 0.184 & -0.129 & 0.031 \\
		-0.001 & 0.000 & -0.001 & 0.002 & 0.099 & 0.066 & -0.727 & -0.675 & -0.030 \\
		-0.002 & -0.000 & 0.000 & 0.001 & -0.361 & 0.150 & -0.645 & 0.653 & 0.062 \\
		-0.000 & 0.000 & 0.000 & -0.000 & 0.040 & -0.035 & 0.011 & -0.053 & 0.997
	\end{pmatrix} \nonumber \\
	& \hspace{2in} \times \frac{1}{\Lambda^2}
	\begin{pmatrix}
		c_{BW} \\
		c_{\phi,1} \\
		c_{DW} \\
		c_{DB} \\
		c_{WWW} \\
		c_W \\
		c_B \\
		c_{WW} \\
		c_{BB}
	\end{pmatrix} =
	\begin{pmatrix}
		-0.004 & \pm & 0.010 \\
		0.062 & \pm & 0.086 \\
		0.022 & \pm & 0.143 \\
		0.628 & \pm & 0.387 \\
		-149.2 & \pm & 120.9 \\
		-17.7 & \pm & 187.5 \\
		589.3 & \pm & 455.1 \\
		-3715 & \pm & 1904 \\
		3902 & \pm & 9964
	\end{pmatrix}\textrm{ TeV$^{-2}$}
	\label{eq:global_bounds}
\end{align}
We find that the tree-level and loop-level bounds are essentially decoupled from each other, as evidenced by the nearly block-diagonal form of the above matrix.  The first four bounds represent bounds on linear combinations of tree-level operators and are very tightly constrained.  The final five are bounds on linear combinations of loop-level operators.  These bounds are weaker than the tree-level bounds by 2 to 3 orders of magnitude or more.  All of the coefficients are consistent with zero at $2\sigma$.

We can also calculate bounds on the loop-level coefficients by first setting the tree-level coefficients to the values (as a function of the loop-level coefficients) that minimize $\chi^2$.  We again write this new $\chi^2$ in matrix form
\begin{equation}
	\chi^2\big|_{\{c_{tree}\}=\{c_{tree}^{min}\}} = \chi^2_{min}
	+ \frac{\sum_{ij}(c_i-\hat{c}_i)M_{ij}(c_j-\hat{c}_j)}{\Lambda^4}
\end{equation}
where $\{c_{tree}\}$ is the set of all tree-level operators, and the sum on $i,j$ runs over all loop-level operators.  We then follow the same procedure as before to arrive at bounds on the loop-level operators
\begin{align}
	\begin{pmatrix}
		-0.913 & -0.218 & 0.145 & -0.312 & 0.011 \\
		-0.156 & 0.961 & 0.184 & -0.129 & 0.031 \\
		-0.099 & -0.066 & 0.727 & 0.675 & 0.030 \\
		0.361 & -0.150 & 0.645 & -0.653 & -0.062 \\
		0.040 & -0.035 & 0.011 & -0.053 & 0.997
	\end{pmatrix} \times \frac{1}{\Lambda^2}
	\begin{pmatrix}
		c_{WWW} \\
		c_W \\
		c_B \\
		c_{WW} \\
		c_{BB}
	\end{pmatrix} =
	\begin{pmatrix}
		-149.2 & \pm & 120.9 \\
		-17.7 & \pm & 187.5 \\
		589.3 & \pm & 455.1 \\
		-3715 & \pm & 1904 \\
		3902 & \pm & 9964
	\end{pmatrix}\textrm{ TeV$^{-2}$}
\label{eq:loop_bounds}
\end{align}
These bounds are identical to the bounds of Eq.~\eqref{eq:global_bounds}, highlighting the fact that the two sets of operators are decoupled in the eigenvector matrix.

We can also find bounds on each loop-level operator separately, by setting all other loop-level operators to zero and letting the relevant tree-level operators float.  This gives us the following bounds
\begin{align}
	\frac{c_{WW}}{\Lambda^2} & = 129.5 \pm 120.8 \textrm{ TeV$^{-2}$} \label{eq:cWW}\\ 
	\frac{c_{BB}}{\Lambda^2} & = 1456 \pm 2225 \textrm{ TeV$^{-2}$} \\
	\frac{c_{WWW}}{\Lambda^2} & = 57.59 \pm 63.09 \textrm{ TeV$^{-2}$} \\
	\frac{c_W}{\Lambda^2} & = 100.6 \pm 181.9 \textrm{ TeV$^{-2}$} \\
	\frac{c_B}{\Lambda^2} & = -123.4 \pm 355.9 \textrm{ TeV$^{-2}$}
\end{align}

\section{Conclusions}
\label{sec:conclusions}

The bounds we have obtained on the loop-level operators are much weaker than the bounds obtained in similar previous analyses \cite{Hagiwara:1992eh,Hagiwara:1993ck,Alam:1997nk}.  These analyses set the renormalized tree-level coefficients to zero rather than letting them float.  This is an unjustified assumption, as these coefficients are renormalized by the one-loop coefficients, as discussed in Section~\ref{sec:tree_renorm}.  Thus the results of these previous analyses are specious, as we discussed in Ref.~\cite{Mebane:2013cra}.  We discuss this further in Appendix~\ref{apdx:ep}.

We found in Ref.~\cite{Mebane:2013cra} that the bounds on the loop-level operators $O_{BB}$ and $O_{WW}$ from precision electroweak physics are much weaker than the bounds from tree-level processes involving the Higgs boson.  The analogous result holds for the other three loop-level operators, $O_{WWW}$, $O_{W}$, and $O_{B}$; they are much more strongly constrained by tree-level processes involving the triple gauge boson vertex \cite{Beringer:1900zz}.  Thus the bounds on the bosonic operators from a one-loop analysis of precision electroweak data cannot compete with the bounds obtained from tree-level processes at high-energy colliders.

\section{Acknowledgements}

This material is based upon work supported in part by the U.~S.\ Department of Energy under Contract No.\ DE-FG02-13ER42001 and the IISN ``Fundamental Interactions'' convention 4.4517.08.

\appendix

\section{Self-Energies}
\label{apdx:se}

The self-energies given below have not been calculated previously in their entirety.  The divergent parts, as well as terms proportional to $m_h^2$ and $\ln m_h^2$ (in the large $m_h$ limit) were calculated in Refs.~\cite{Hagiwara:1992eh,Hagiwara:1993ck,Alam:1997nk}, and we have confirmed these previous calculations.

\subsection{Tree-Level Contributions}
\label{apdx:tree_se}

\begin{align}
	\Pi_{WW} \quad & = \quad -\frac{c_{DW}}{\Lambda^2} 2g^2 q^4 \\
	\Pi_{ZZ} \quad & = \quad -\frac{c_{BW}}{\Lambda^2} 2m_W^2 s^2 q^2 +
	\frac{c_{\phi,1}}{\Lambda^2} \frac{v^2}{2}m_Z^2 - \frac{c_{DW}}{\Lambda^2} 2g^2 c^2 q^4
	- \frac{c_{DB}}{\Lambda^2} 2g^2 \frac{s^4}{c^2} q^4 \\
	\Pi_{\gamma\gamma} \quad & = \quad \frac{c_{BW}}{\Lambda^2} 2m_W^2 s^2 q^2
	- \frac{c_{DW}}{\Lambda^2} 2g^2 s^2 q^4 - \frac{c_{DB}}{\Lambda^2} 2g^2 s^2 q^4 \\
	\Pi_{\gamma Z} \quad & = \quad \frac{c_{BW}}{\Lambda^2} m_W^2 \frac{s}{c} (c^2-s^2)q^2
	- \frac{c_{DW}}{\Lambda^2} 2g^2 s c q^4 + \frac{c_{DB}}{\Lambda^2} 2g^2 \frac{s^3}{c} q^4
\end{align}

\subsection{One-Loop Contributions}
\label{apdx:loop_se}

The expressions below are given in terms of scalar integral functions $A_0$, $B_0$, and $C_0$. Expressions for these functions are given in Appendix D of Ref.~\cite{Passarino:1978jh}

\noindent $\mathcal{O}_{BB}$:
\nopagebreak
\vspace{14pt}
\nopagebreak
\begin{align}
	\overline{\Pi}_{WW} \quad & = \quad \frac{c_{BB}}{\Lambda^2} \frac{1}{16\pi^2} \frac{g^2 m_Z^4 s^4}{m_h^2}\left(2m_Z^2-3 A_0\left(m_Z^2\right)\right) \\
	\overline{\Pi}_{ZZ} \quad & = \quad \frac{c_{BB}}{\Lambda^2} \frac{1}{16\pi^2} \frac{g^2 s^4}{m_h^2 c^2}\left[2 m_h^2 m_Z^2\left(q^2-m_h^2 + m_Z^2\right)B_0\left(q^2,m_h^2,m_Z^2\right)\right. \\
	& \qquad - m_Z^2 \left(3 q^2  + 2 m_h^2 + 3m_Z^2\right)A_0\left(m_Z^2\right) + m_h^2 \left(2m_Z^2 - q^2\right) A_0\left(m_h^2\right) \nonumber \\
	& \qquad \left. - 6 m_W^2 q^2 A_0\left(m_W^2\right) + 4m_W^4 q^2 + 2 m_Z^4 q^2 + 2m_Z^6\right] \nonumber \\
	\overline{\Pi}_{\gamma\gamma} \quad & = \quad -\frac{c_{BB}}{\Lambda^2} \frac{1}{16\pi^2} \frac{g^2 q^2 s^2}{m_h^2}\left[m_h^2 A_0\left(m_h^2\right) + 3m_Z^2 A_0\left(m_Z^2\right) \right. \\
	& \qquad \left. + 6m_W^2 A_0\left(m_W^2\right) -4m_W^4 - 2m_Z^4\right] \nonumber \\
	\overline{\Pi}_{\gamma Z} \quad & = \quad \frac{c_{BB}}{\Lambda^2} \frac{1}{16\pi^2} \frac{g^2 s^3}{m_h^2 c} \left[m_h^2 m_Z^2\left(m_h^2 - m_Z^2 - q^2\right)B_0\left(q^2,m_h^2,m_Z^2\right)\right. \\
	& \qquad + m_Z^2\left(m_h^2 + 3q^2\right) A_0\left(m_Z^2\right) - m_h^2\left(m_Z^2-q^2\right) A_0\left(m_h^2\right) \nonumber \\
	& \qquad \left. + 6 m_W^2 q^2 A_0\left(m_W^2\right) - 4m_W^4 q^2 - 2m_Z^4 q^2\right] \nonumber
\end{align}

\vspace{20pt}
\noindent $\mathcal{O}_{WW}$:
\nopagebreak
\vspace{14pt}
\nopagebreak
\begin{align}
	\overline{\Pi}_{WW} \quad & = \quad \frac{c_{WW}}{\Lambda^2} \frac{1}{16\pi^2} \frac{g^2}{m_h^2} \left[2m_h^2 m_W^2 \left(q^2-m_h^2+m_W^2\right) B_0\left(q^2,m_h^2,m_W^2\right)\right. \\
	& \qquad - 2m_W^2\left(3q^2 + m_h^2 + 3m_W^2\right) A_0\left(m_W^2\right) + m_h^2\left(2m_W^2 - q^2\right)A_0\left(m_h^2\right) \nonumber \\
	& \qquad \left. - 3\left(m_W^4 + m_Z^2 q^2\right) A_0\left(m_Z^2\right) + 4m_W^4 q^2 + 2m_Z^4 q^2 + 4m_W^6 + 2m_W^4 m_Z^2\right] \nonumber \\
	\overline{\Pi}_{ZZ} \quad & = \quad \frac{c_{WW}}{\Lambda^2} \frac{1}{16\pi^2} \frac{g^2 c^2}{m_h^2} \left[2m_h^2 m_Z^2\left(q^2 - m_h^2 + m_Z^2\right) B_0\left(q^2,m_h^2,m_Z^2\right) \right. \\
	& \qquad - m_Z^2\left(3q^2 + 2m_h^2 + 3m_Z^2\right) A_0\left(m_Z^2\right) + m_h^2 \left(2m_Z^2 - q^2\right) A_0\left(m_h^2\right) \nonumber \\
	& \qquad \left. - 6\left(m_Z^4 + m_W^2 q^2\right) A_0\left(m_W^2\right) + 4 m_W^4 q^2 + 2 m_Z^4 q^2 + 4m_W^2 m_Z^4 + 2 m_Z^6\right] \nonumber \\
	\overline{\Pi}_{\gamma\gamma} \quad & = \quad -\frac{c_{WW}}{\Lambda^2} \frac{1}{16\pi^2} \frac{g^2 q^2 s^2}{m_h^2} \left[m_h^2 A_0\left(m_h^2\right) + 3 m_Z^2 A_0\left(m_Z^2\right) \right. \\
	& \qquad \left. + 6 m_W^2 A_0\left(m_W^2\right) - 4m_W^4 - 2m_Z^4\right] \nonumber \\
	\overline{\Pi}_{\gamma Z} \quad & = \quad -\frac{c_{WW}}{\Lambda^2} \frac{1}{16\pi^2} \frac{g^2 s c}{m_h^2} \left[m_h^2 m_Z^2\left(m_h^2 - m_Z^2 - q^2\right) B_0\left(q^2,m_h^2,m_Z^2\right) \right. \\
	& \qquad + m_Z^2\left(m_h^2 + 3q^2\right)A_0\left(m_Z^2\right) - m_h^2 \left(m_Z^2 - q^2\right) A_0\left(m_h^2\right) \nonumber \\
	& \qquad \left. + 6 m_W^2 q^2 A_0\left(m_W^2\right) - 4 m_W^4 q^2 - 2 m_Z^4 q^2\right] \nonumber
\end{align}

\vspace{20pt}
\noindent $\mathcal{O}_{B}$:
\nopagebreak
\vspace{14pt}
\nopagebreak
\begin{align}
	\overline{\Pi}_{WW} \quad & = \quad \frac{c_B}{\Lambda^2} \frac{1}{16\pi^2}
	\frac{g^2 m_W^2 s^2}{36q^2 c^2} \left[36m_W^2(q^4-m_W^4)C_0(0,0,q^2,m_W^2,0,m_Z^2) \right. \\
	& \qquad \left. - 6c^2(m_W^2 - q^2)^2 B_0(q^2,0,m_W^2)
	+ 3((-2s^6+19s^4-30s^2+12)m_Z^4 \right. \nonumber \\
	& \qquad \left. - 2(2s^4+7s^2-7)m_Z^2 q^2 + (5+2c^2)q^4)B_0(q^2,m_W^2,m_Z^2)
	- 3((1+3c^2)m_Z^2 + 5q^2)A_0(m_W^2) \right. \nonumber \\
	& \qquad \left. + 3(s^2-c^2)(m_Z^2+5m_W^2 - 5q^2)A_0(m_Z^2)
	+ 2q^2(3s^2 m_Z^2 - q^2)\right] \nonumber \\
	\overline{\Pi}_{ZZ} \quad & = \quad \frac{c_B}{\Lambda^2} \frac{1}{16\pi^2} \frac{g^2 s^2}{36q^2 c^2}
	\left[72m_W^4 q^2(c^2 q^2 - m_W^2)C_0(0,0,q^2,m_W^2,0,m_W^2) \right. \\
	& \qquad \left. + 3(-m_Z^2(m_Z^2-m_h^2)^2 + (m_Z^4-8m_h^2 m_Z^2 -m_h^4)q^2
	+ (m_Z^2 + 2m_h^2)q^4 - q^6)B_0(q^2,m_h^2,m_Z^2) \right. \nonumber \\
	& \qquad \left. + 3q^2(24m_W^4 + 8(c^2-s^2)m_W^2 q^2 + (c^2-s^2)q^4)
	B_0(q^2,m_W^2,m_W^2) \right. \nonumber \\
	& \qquad \left. + 3(m_h^2m_Z^2 - m_Z^4 + m_h^2 q^2 + q^4)A_0(m_h^2)
	+ 3(m_Z^4 - m_h^2 m_Z^2 - (10m_Z^2 + m_h^2)q^2 + q^4)A_0(m_Z^2) \right. \nonumber \\
	& \qquad \left. + 6q^2(-12m_W^2 + (10c^2+1)q^2)A_0(m_W^2) \right. \nonumber \\
	& \qquad \left. + 2q^2(3m_Z^4 + 3m_h^2 m_Z^2 + (3m_h^2 - 2(6s^4-9s^2+2)m_Z^2)q^2 - 2s^2 q^4)
	\right] \nonumber \\
	\overline{\Pi}_{\gamma\gamma} \quad & = \quad -\frac{c_B}{\Lambda^2} \frac{1}{16\pi^2}
	\frac{g^2 q^2 s^2}{18} \left[36m_W^4 C_0(0,0,q^2,m_W^2,0,m_W^2) \right. \\
	& \qquad \left. + 3(8m_W^2 + q^2)B_0(q^2,m_W^2,m_W^2) + 30A_0(m_W^2) + 2(q^2 - 6m_W^2)
	\right] \nonumber \\
	\overline{\Pi}_{\gamma Z} \quad & = \quad \frac{c_B}{\Lambda^2} \frac{1}{16\pi^2} \frac{g^2 s}{72c}
	\left[72m_W^4(m_W^2 + (s^2-c^2)q^2)C_0(0,0,q^2,m_W^2,0,m_W^2) \right. \\
	& \qquad \left. + 3(m_h^4 + 4m_h^2 m_Z^2 - 5m_Z^4 - (2m_h^2 - 4m_Z^2)q^2 + q^4)
	B_0(q^2,m_h^2,m_Z^2) \right. \nonumber \\
	& \qquad \left. - 3(24m_W^4 + 8(c^2-3s^2)m_W^2 q^2 + (c^2-3s^2)q^4)B_0(q^2,m_W^2,m_W^2)
	- 3(5m_Z^2 + m_h^2 + q^2)A_0(m_h^2) \right. \nonumber \\
	& \qquad \left. + 3(5m_Z^2 + m_h^2 - q^2)A_0(m_Z^2)
	+ 6(12m_W^2 + (9s^2-11c^2)q^2)A_0(m_W^2) \right. \nonumber \\
	& \qquad \left. + 2q^2(3(8s^4-10s^2+1)m_Z^2 - 3m_h^2 + 4s^2 q^2)\right] \nonumber
\end{align}

\vspace{20pt}
\noindent $\mathcal{O}_{W}$:
\nopagebreak
\vspace{14pt}
\nopagebreak
\begin{align}
	\overline{\Pi}_{WW} \quad & = \quad \frac{c_W}{\Lambda^2} \frac{1}{16\pi^2} \frac{g^2}{12q^2} \left[
	12m_W^2\left(m_W^4(m_W^2 + m_Z^2 + 2 q^2) - \left(3m_W^2 + m_Z^2\right)q^4\right)
	C_0(0,0,q^2,m_W^2,0,m_Z^2)\right. \\
	& \qquad \left. - (m_W^2(m_h^2 - m_W^2)^2
	+ q^2(m_h^4 + 8m_h^2 m_W^2 - m_W^4) - q^4(2m_h^2+m_W^2) + q^6) B_0(q^2,m_h^2,m_W^2)
	\right. \nonumber \\
	& \qquad \left. - 2s^2 m_W^2(m_W^2 - q^2)^2 B_0(q^2,0,m_W^2)
	+ \left(2(s^6-10s^4+24s^2-12)m_W^2 m_Z^4 \right.\right. \nonumber \\
	& \qquad \left.\left. - (4s^6+45s^4-106s^2+52)m_Z^4 q^2
	- 2(s^4-10s^2+11)m_Z^2 q^4 - q^6\right) B_0(q^2,m_W^2,m_Z^2) \right. \nonumber \\
	& \qquad \left. + (m_h^2(m_W^2 + q^2) - m_W^4 + q^4)A_0(m_h^2) \right. \nonumber \\
	& \qquad \left. - \left(m_h^2 m_W^2 - 6m_W^2 m_Z^2 - 7m_W^4
	+ \left(m_h^2 - 5m_Z^2 - 11m_W^2\right)q^2 + 22q^4\right)	A_0(m_W^2) \right. \nonumber \\
	& \qquad \left. + \left(2(5s^4-14s^2+6)m_W^2 m_Z^2
	+ (14s^4-45s^2+26)m_Z^2 q^2 - (23c^2-s^2)q^4\right)A_0(m_Z^2)
	\right. \nonumber \\
	& \qquad \left. + 2q^2\left(m_h^2 m_W^2 - 12m_W^2 m_Z^2 - 7m_W^4 + (m_h^2 + m_W^2 + m_Z^2)q^2 - \frac{2}{3}q^4\right)
	\right] \nonumber	\\
	\overline{\Pi}_{ZZ} \quad & = \quad \frac{c_W}{\Lambda^2} \frac{1}{16\pi^2} \frac{g^2}{12 q^2} \left[
	24m_W^2 m_Z^2(m_W^4 + m_W^2 q^2 - c^2(1+c^2)q^4)C_0(0,0,q^2,m_W^2,0,m_W^2) \right. \\
	& \qquad \left.
	+ (-m_Z^2(m_Z^2 - m_h^2)^2 - (m_h^4 + 8m_h^2 m_Z^2 - m_Z^4)q^2 + (2m_h^2 + m_Z^2)q^4 - q^6)
	B_0(q^2,m_h^2,m_Z^2) \right. \nonumber \\
	& \qquad \left.
	- (24m_W^4 m_Z^2 + 4(m_W^2 m_Z^2 + 12m_W^4)q^2 + 2(8s^4-24s^2+11)m_Z^2 q^4
	\right. \nonumber \\
	& \qquad \left. + (c^2-s^2)q^6)	B_0(q^2,m_W^2,m_W^2)
	+ (m_h^2 m_Z^2 - m_Z^4 + m_h^2 q^2 + q^4)A_0(m_h^2) \right. \nonumber \\
	& \qquad \left. + (m_Z^4 - m_h^2 m_Z^2 - (10m_Z^2 + m_h^2)q^2 + q^4)A_0(m_Z^2) \right. \nonumber \\
	& \qquad \left. + 2(12m_W^2 m_Z^2 + 2(m_Z^2 + 12m_W^2)q^2 - (13 + 10c^2)q^4)A_0(m_W^2)
	\right. \nonumber \\
	& \qquad \left. + q^2\left(2m_h^2 m_Z^2 - 2(24s^4-44s^2+19)m_Z^4 + (2m_h^2 + 4(c^2-s^2)m_W^2)q^2
	- \frac{4c^2}{3}q^4\right)\right] \nonumber	\\
	\overline{\Pi}_{\gamma\gamma} \quad & = \quad -\frac{c_W}{\Lambda^2} \frac{1}{16\pi^2}
	 \frac{g^2 q^2 s^2}{6} \left[
	12m_W^4 C_0(0,0,q^2,m_W^2,0,m_W^2) + (8m_W^2+q^2)B_0(q^2,m_W^2,m_W^2) \right. \\
	& \qquad \left. + 10A_0(m_W^2) - 4m_W^2 + \frac{2q^2}{3}\right] \nonumber \\
	\overline{\Pi}_{\gamma Z} \quad & = \quad \frac{c_W}{\Lambda^2} \frac{1}{16\pi^2}
	\frac{g^2 s}{72 c} \left[-72(1+2c^2)m_W^4 q^2 C_0(0,0,q^2,m_W^2,0,m_W^2) \right. \\
	& \qquad \left. + 3((m_Z^2-m_h^2)(5m_Z^2+m_h^2) + (2m_h^2 - 4m_Z^2)q^2 - q^4)B_0(q^2,m_h^2,m_Z^2)
	\right. \nonumber \\
	& \qquad \left. - 3(48m_W^4 + 16(1+2c^2)m_W^2 q^2 + (3c^2-s^2)q^4)B_0(q^2,m_W^2,m_W^2)
	\right. \nonumber \\
	& \qquad \left. + 3(m_h^2 + 5m_Z^2 + q^2)A_0(m_h^2)
	- 3(m_h^2 + 5m_Z^2 - q^2)A_0(m_Z^2) \right. \nonumber \\
	& \qquad \left. + 6(24m_W^2 - (13+20c^2)q^2)A_0(m_W^2) \right. \nonumber \\
	& \qquad \left. - 2(72m_W^4 - 3((8s^4-14s^2+7)m_Z^2 + m_h^2)q^2 + 4c^2 q^4)\right] \nonumber
\end{align}

\vspace{20pt}
\noindent $\mathcal{O}_{WWW}$:
\nopagebreak
\vspace{14pt}
\nopagebreak
\begin{align}
	\overline{\Pi}_{WW} \quad & = \quad \frac{c_{WWW}}{\Lambda^2} \frac{1}{16\pi^2} 3g^4 \left[
	2m_W^4(m_W^2-q^2) C_0\left(0,0,q^2,m_W^2,0,m_Z^2\right)\right. \\
	& \qquad \left. + m_Z^2\left((s^2-2c^2)m_W^2 - c^2 q^2\right) B_0\left(q^2,m_W^2,m_Z^2\right)
	- \left((s^2-c^2)m_W^2 + 2c^2 q^2\right) A_0\left(m_Z^2\right)\right. \nonumber \\
	& \qquad \left. + \left(m_W^2 - 2q^2\right)A_0(m_W^2)
	+ \frac{3}{2} m_W^2 q^2 + \frac{q^4}{6}\right] \nonumber \\
	\overline{\Pi}_{ZZ} \quad & = \quad \frac{c_{WWW}}{\Lambda^2} \frac{1}{16\pi^2} 3g^4 \left[
	2m_W^4(m_W^2 - c^2 q^2) C_0\left(0,0,q^2,m_W^2,0,m_W^2\right) \right. \\
	& \qquad \left . - m_W^2(2m_W^2 + (c^2-s^2)q^2) B_0\left(q^2,m_W^2,m_W^2\right)
	+ 2(m_W^2 - 2c^2 q^2)A_0\left(m_W^2\right) \right. \nonumber \\
	& \qquad \left. + \frac{1}{2} m_W^2 (3c^2 - s^2) q^2
	+ \frac{1}{6}c^2 q^4\right] \nonumber \\
	\overline{\Pi}_{\gamma\gamma} \quad & = \quad -\frac{c_{WWW}}{\Lambda^2} \frac{1}{16\pi^2}
	6g^4 q^2 s^2 \left[ m_W^4 C_0\left(0,0,q^2,m_W^2,0,m_W^2\right)
	+ m_W^2 B_0\left(q^2,m_W^2,m_W^2\right) \right. \\
	& \qquad \left. + 2A_0(m_W^2) - m_W^2 - \frac{q^2}{12}\right] \nonumber \\
	\overline{\Pi}_{\gamma Z} \quad & = \quad \frac{c_{WWW}}{\Lambda^2} \frac{1}{16\pi^2} \frac{3g^4}{2}
	\frac{s}{c} \left[ 2m_W^4(m_W^2 - 2c^2 q^2)C_0(0,0,q^2,m_W^2,0,m_W^2) \right. \\
	& \qquad \left. - m_W^2(2m_W^2 + (3c^2-s^2)q^2)B_0(q^2,m_W^2,m_W^2)
	+ 2(m_W^2 - 4c^2q^2)A_0(m_W^2) \right. \nonumber \\
	& \qquad \left. + \frac{1}{2}m_W^2 (7c^2 - s^2)q^2 + \frac{c^2 q^4}{3}\right] \nonumber
\end{align}

\section{Comparison with previous bounds}
\label{apdx:ep}

Here we discuss in detail why the bounds we obtain on the coefficients of dimension-six operators are so much weaker than the bounds obtained in previous calculations \cite{Hagiwara:1992eh,Hagiwara:1993ck,Alam:1997nk}.  We focus on the coefficient $c_{WW}$, but the story is similar for all of the one-loop coefficients.

\begin{figure}
\includegraphics[width=5in]{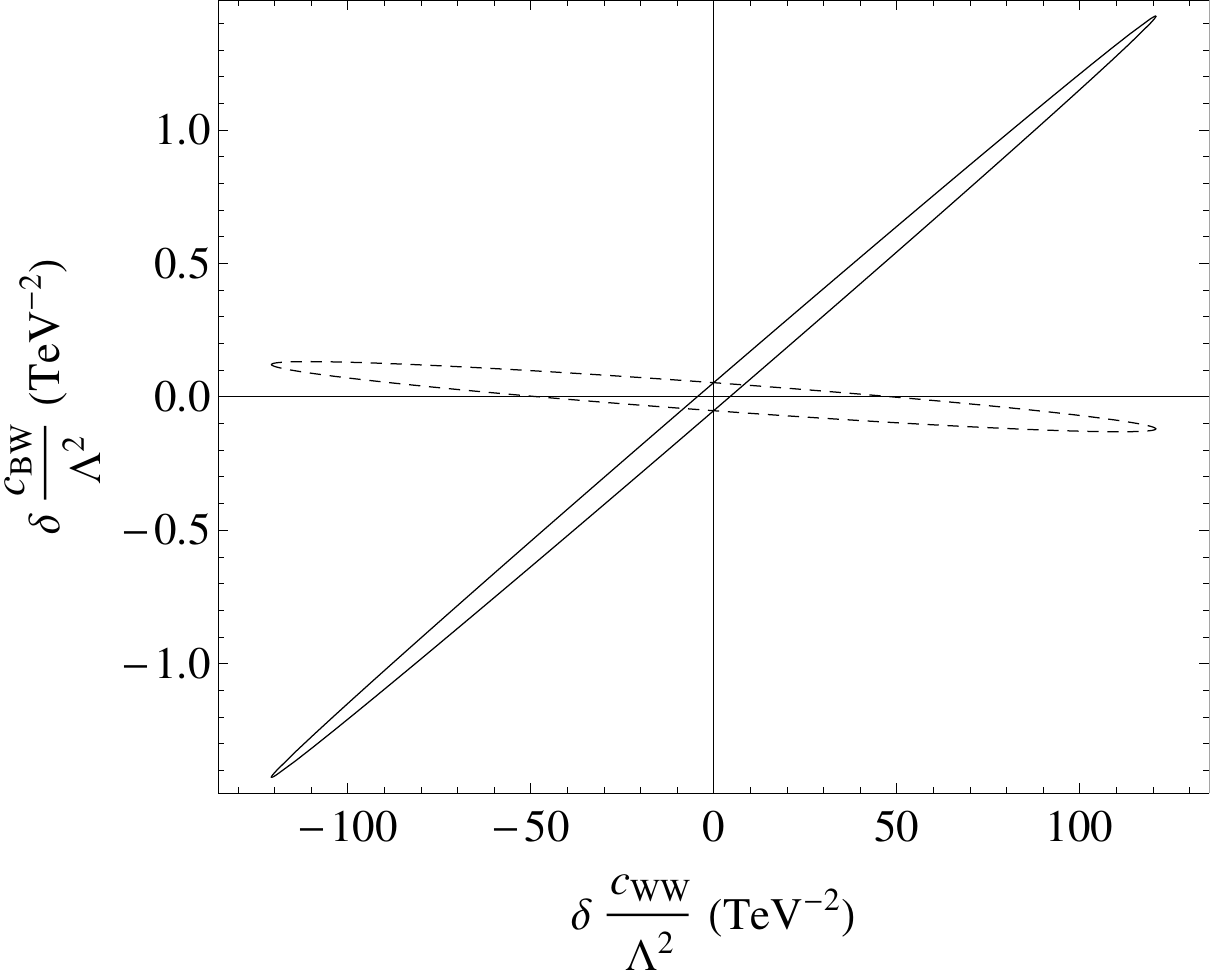}
\caption{Two-parameter fit to precision electroweak data.  The tree-level parameter $c_{BW}$ and the one-loop parameter $c_{WW}$ are centered at their best-fit values and allowed to float.  Dashed ellipse: Renormalization scale of $M_Z$.  Solid ellipse: Renormalization scale of 1 TeV.}\label{fig:ep}
\end{figure}

We show in Fig.~\ref{fig:ep} a two-parameter fit of the tree-level coefficient $c_{BW}$ and the one-loop coefficient $c_{WW}$ to the precision electroweak data.  We have centered the coefficients at their best-fit values.  The dashed ellipse corresponds to a renormalization scale of $M_Z$, which is the appropriate scale.  If both coefficients are allowed to float, the bound on $c_{WW}$ is given by the full extent of the ellipse in the horizontal direction, $\pm 120.8$ TeV$^{-2}$ ({\it cf.} Eq.~(\ref{eq:cWW})).  If the tree-level coefficient $c_{BW}$ is fixed to its central value, however, the bound on $c_{WW}$ is given by the intercept of the ellipse with the horizontal axis, $\pm 48.3$ TeV$^{-2}$.  This partially explains the tighter bounds obtained in Refs.~\cite{Hagiwara:1992eh,Hagiwara:1993ck,Alam:1997nk}.

There is another factor, however, and that is the choice of renormalization scale.  In Refs.~\cite{Hagiwara:1992eh,Hagiwara:1993ck,Alam:1997nk}, the renormalization scale was chosen to be $\Lambda = 1$ TeV.  This has the effect of enhancing the one-loop calculations of Refs.~\cite{Hagiwara:1992eh,Hagiwara:1993ck,Alam:1997nk}, which contain
terms proportional to $\ln\Lambda^2$.  We show in Fig.~\ref{fig:ep} a fit with the renormalization scale set to 1 TeV (solid ellipse).  If both coefficients are allowed to float, the bound on $c_{WW}$ is the same as before, which demonstrates that our bound is independent of the renormalization scale.  If the tree-level coefficient $c_{BW}$ is fixed to its central value, however, the bound on $c_{WW}$ is given by the intercept of the solid ellipse with the horizontal axis, $\pm 4.45$ TeV$^{-2}$.  This is a much tighter bound than the true bound of $\pm 120.8$ TeV$^{-2}$.

\end{document}